\documentclass[11pt]{article}

\usepackage[T1]{fontenc}
\usepackage[utf8]{inputenc}
\usepackage{lmodern}
\usepackage{tipa}
\usepackage{microtype}
\usepackage[margin=1in]{geometry}
\usepackage{setspace}
\usepackage{booktabs}
\usepackage{tabularx}
\usepackage{array}
\usepackage{enumitem}
\usepackage{graphicx}
\usepackage{tikz}
\usetikzlibrary{positioning,arrows.meta,fit,calc}
\usepackage{natbib}
\usepackage[hidelinks]{hyperref}
\usepackage[nameinlink,noabbrev]{cleveref}
\usepackage{xcolor}
\usepackage{ragged2e}
\usepackage{caption}
\usepackage{titlesec}
\usepackage{fancyhdr}

\hypersetup{
  pdftitle={Judgment-Centred Software Engineering Education: A Post-Hype Review and Framework for AI-Augmented Learning},
  pdfauthor={Qusay H. Mahmoud},
  pdfsubject={Software engineering education, generative AI, agentic software engineering},
  pdfkeywords={generative AI, software engineering education, agentic software engineering, assessment, comprehension debt, human-AI collaboration}
}

\setlist{nosep,leftmargin=*}

\titleformat{\section}{\large\bfseries}{\thesection}{0.6em}{}
\titleformat{\subsection}{\normalsize\bfseries}{\thesubsection}{0.6em}{}
\newcommand{\obligation}[1]{\textsc{#1}}

\newcolumntype{Y}{>{\RaggedRight\arraybackslash}X}
\newcolumntype{P}[1]{>{\RaggedRight\arraybackslash}p{#1}}

\title{\bfseries Judgment-Centred Software Engineering Education:\\
\large A Post-Hype Review and Framework for AI-Augmented Learning}
\author{
Qusay H. Mahmoud\\
Faculty of Engineering and Applied Science\\
Ontario Tech University\\
Oshawa, Ontario, Canada\\
\texttt{qusay.mahmoud@ontariotechu.ca}
}
\date{September 2026}

\begin{document}
\frenchspacing
\maketitle

\begin{abstract}
Generative artificial intelligence has moved from a disruptive novelty to a recurring part of software-development and computing-education workflows, while software agents are beginning to act across repositories, command lines, browsers, tests, and other tools.  The educational problem is therefore no longer captured by the question of whether students should be allowed to generate code.  It is whether software-engineering (SE) programs can preserve and assess human understanding while preparing students to work responsibly with increasingly capable AI systems.  This paper presents a structured integrative review of research and practice from 2023 through 23 September 2026, supplemented by established work on AI literacy, technical debt, and human--AI collaboration.  Recent systematic reviews and meta-analytic evidence strengthen a conditional conclusion: GenAI can improve access to explanations, feedback, practice, and short-term task completion, but effects on externally assessed learning are highly heterogeneous and depend on prior knowledge, scaffolding, verification, task design, and the form of outcome being measured.  We also examine the human actors around those outcomes: how students use AI as a help-seeking resource and negotiate ownership of AI-assisted work, and how faculty face changed assessment evidence, workload, and policy demands.  In SE, the stakes widen because generated artifacts enter shared repositories, persist across iterations, interact with architecture and security constraints, and may affect clients and deployed systems.  We therefore organize the evidence around \emph{accountable engineering judgment}.  We credit and extend the emerging concept of \emph{comprehension debt}: the deferred learning and maintenance cost that appears when AI-assisted production outpaces a learner's or team's ability to explain, test, modify, and justify the resulting software.  We then refine the AI-Augmented Software Engineering Education (AASEE) framework into five non-linear integration levels, from tool awareness to agentic and AI systems engineering.  Across all levels, four evidence-of-learning obligations---\obligation{explain}, \obligation{verify}, \obligation{modify}, and \obligation{account}---make retained human competence visible.  The resulting design logic is neither prohibitionist nor technology-first: learning objectives determine delegation, and delegation is paired with evidence, governance, and recovery mechanisms commensurate with its consequences.
\end{abstract}


\section{Introduction}

The release of broadly accessible large language models (LLMs) changed a basic assumption of computing education: a plausible software artifact no longer reliably reveals the reasoning that produced it.  A student can obtain code, tests, explanations, documentation, user stories, architectural alternatives, and project plans with little visible trace of which parts were independently understood.  Early computing-education work correctly treated this as a disruption to programming pedagogy and assessment \citep{becker2023,denny2024}.  SE-focused position and review work quickly broadened the debate toward deliberate integration, professional formation, and lifecycle responsibility \citep{bull2024,tenbergen2024,sengul2024}.  Subsequent work has shown that the disruption is not reducible to academic misconduct.  Students use GenAI for ordinary help-seeking, explanation, debugging, ideation, and documentation; educators are simultaneously trying to preserve learning, redesign assessment, and prepare students for AI-mediated professional practice \citep{zastudil2023,shoufan2023,keuning2024,kizilcec2024,yabaku2024}.

By 2026 the evidence base is materially stronger than it was during the first wave of debate.  Broad syntheses now cover computing and programming education at scale \citep{prather2025,agbo2025,kumar2026,adejumo2026,katona2026,contreras2026}, code comprehension has received a dedicated systematic review \citep{qiao2026}, and a recent STEM meta-analysis demonstrates why simple claims that GenAI ``improves learning'' are unsafe: a positive conventional pooled effect coexists with extreme between-study heterogeneity, and the overall effect becomes indistinguishable from zero after a publication-bias correction in that analysis \citep{boolzen2026}.  The mature question is therefore not whether GenAI is good or bad for learning.  It is \emph{which forms of AI-mediated activity develop which forms of competence, for which learners, under which conditions}.

That question is especially important in software engineering.  Programming education often centres on algorithmic reasoning, syntax, debugging, and small-scale code.  SE education includes these foundations but extends to requirements, architecture, testing strategy, configuration and release practices, security, maintenance, teamwork, client communication, professional ethics, and accountability across a software lifecycle.  Generated code can be locally correct while being architecturally inappropriate, insecure, unmaintainable, inconsistent with stakeholder needs, or poorly understood by the team that inherits it.  A final artifact is thus an increasingly weak proxy for professional competence.

A second shift makes the problem more urgent.  AI support is moving from conversational assistance and inline completion toward \emph{agentic} systems that can inspect repositories, edit multiple files, invoke command-line tools, execute tests, browse documentation, and iteratively pursue goals.  SWE-bench helped move evaluation from isolated code snippets to real repository issues requiring multi-file and execution-environment reasoning \citep{jimenez2024}; OpenHands demonstrates generalist software agents that write code, use a command line, browse, and operate in sandboxed environments \citep{wang2025}.  These papers establish technical capability, not educational benefit.  Studies of code-generation assistants already show that programmers alternate between acceleration and exploration, accepting, modifying, checking, and rejecting suggestions rather than interacting with AI in a single uniform mode \citep{barke2023}.  But they change what authentic preparation can mean: graduates increasingly need to supervise delegated engineering work rather than merely prompt a chatbot.

This paper takes a \emph{post-hype} perspective.  Post-hype does not mean that AI is settled, unimportant, or safe.  It means that nearly four years of experimentation now permit sharper distinctions among productivity, performance, understanding, transfer, and professional readiness.  The central argument is that SE education should move from a production-centred model to a \emph{judgment-centred} model.  Students still need programming, testing, design, maintenance, teamwork, requirements, architecture, security, and ethics.  AI makes these fundamentals more important because they are the basis on which generated work is evaluated.  Judgment is layered: students judge when to seek and delegate help and whether output is trustworthy; instructors judge what counts as credible evidence of learning; teams judge whether delegated changes are safe to integrate; and programs judge where AI should be restricted, permitted, required, or governed.

The paper makes four contributions.  First, it updates and synthesizes the 2023--2026 evidence with explicit attention to evidence strength, student and faculty perspectives, and the limits of generalizing short-term computing-education results to lifecycle-wide SE practice.  Second, it treats the emerging shift from copilots to agents as an educational change in supervision, permissions, evaluation, provenance, recovery, and accountability.  Third, building on Ahmad's empirical account of comprehension debt \citep{ahmad2026}, it operationalizes the construct as a curriculum and assessment lens: a gap between what a learner or team can produce with AI and what they can later explain, test, modify, and justify.  Fourth, it refines the AI-Augmented Software Engineering Education (AASEE) framework into five integration levels crossed by four evidence obligations---\obligation{explain}, \obligation{verify}, \obligation{modify}, and \obligation{account}.  The result is a design framework for deciding not simply whether AI is present, but what work is delegated and what evidence shows that human responsibility remains intact.

\section{Review scope and method}

\subsection{A structured integrative review}

This paper is a structured integrative review rather than a registered systematic review or statistical meta-analysis.  That choice reflects both the heterogeneity of the literature and the purpose of the paper.  The relevant evidence includes controlled and quasi-experimental studies, surveys, observational studies, design-science prototypes, course and capstone reports, systematic reviews, conceptual SE papers, and emerging work on software agents.  These sources answer different questions and should not be pooled as if they estimated a common treatment effect.

The main search window for this revision was January 2023 through 23 September 2026.  Sources were identified and re-verified using ACM Digital Library, IEEE Xplore, SpringerLink, ScienceDirect, Frontiers, Google Scholar, arXiv, official conference or institutional publication pages, and backward/forward citation chaining from high-coverage reviews.  Search families combined terms for generative AI, large language models, ChatGPT, Copilot, AI agents, and agentic software engineering with software engineering education, computing education, programming education, assessment, academic integrity, tutoring, feedback, code comprehension, software testing, capstone projects, human--AI collaboration, and curriculum design.

Publications were retained when they contributed directly to at least one of four concerns: (1) learning outcomes or student/instructor use of GenAI in programming, computing, or SE education; (2) a pedagogical mechanism such as guarded tutoring, feedback, assessment redesign, or structured AI use; (3) an SE lifecycle context with educational implications, such as testing or client-facing capstones; or (4) technical or conceptual developments in software agents that materially change what authentic SE supervision entails.  General AI-in-education work without computing relevance, product marketing, and purely promotional commentary were excluded from the evidentiary core.  Older sources are used only when they provide a stable conceptual anchor, such as AI literacy \citep{long2020}, technical debt \citep{kruchten2012}, or calibrated human--AI reliance \citep{bansal2021}.

The original working review did not retain database-specific hit counts and screening decisions in a form that would support a defensible PRISMA flow.  This revision therefore does not invent such counts or claim exhaustive retrieval.  Instead, it improves verifiability by (a) stating the search window and search families, (b) separating evidence types, (c) preferring version-of-record publications when available, (d) distinguishing educational evidence from technical capability evidence, and (e) using recent systematic reviews as coverage anchors.  The resulting claims should be interpreted as an evidence-informed integrative synthesis, not as prevalence estimates.

\subsection{Evidence weighting}

The synthesis gives the greatest weight to systematic reviews, meta-analytic evidence, and peer-reviewed empirical studies when making claims about student outcomes.  Tool and design studies are used primarily to identify pedagogical mechanisms.  Experience reports and qualitative or small-sample observational studies contribute contextual detail but do not establish broad causal effects.  Preprints are retained only where they address fast-moving 2026 developments with unusually direct relevance, and are labelled as emerging evidence or conceptual proposals.  Repository-agent papers are used to characterize technical capabilities, not educational effectiveness.  Recent conceptual syntheses of the emerging `agentic engineer' are treated as future-facing curriculum proposals rather than outcome evidence \citep{alenezi2026}.

This weighting is important because ``GenAI use'' is not a single intervention.  A guarded hint system, unrestricted code generation, structured prompt training, AI-assisted unit testing, and a repository agent all delegate different work and create different learning risks.  Likewise, outcome measures differ: speed, assignment score, self-efficacy, conceptual knowledge, independent performance, maintenance ability, and professional judgment are not interchangeable.

Table~\ref{tab:evidence} summarizes how the different evidence types are used in this synthesis and the principal interpretive limit of each.

\begin{table}[t]
\centering
\caption{How evidence types are used in this synthesis.}
\label{tab:evidence}
\small
\begin{tabularx}{\textwidth}{P{0.20\textwidth}Y Y}
\toprule
\textbf{Evidence type} & \textbf{Strongest use} & \textbf{Interpretive limit} \\
\midrule
Systematic reviews / meta-analysis & Cross-study patterns, heterogeneity, recurrent risks, research gaps & Inherit heterogeneity and reporting limitations of primary studies \\
Controlled or quasi-experimental studies & Effects of a specific intervention under specified conditions & Often short-term; tool and task specific \\
Surveys / perception studies & Adoption, perceived usefulness, policy ambiguity, attitudes & Perceptions are not durable learning outcomes \\
Observational / qualitative / capstone studies & Workflow, team, lifecycle, and stakeholder context & Limited generalizability; usually small or situated samples \\
Design/tool studies & Mechanisms for scaffolding, feedback, guardrails, and practice & Technical quality or usability does not itself establish learning \\
Agent benchmarks/platforms & What delegated software work is technically possible & Do not establish educational benefit or appropriate classroom use \\
Conceptual/preprint work & Emerging vocabulary, design propositions, future-facing questions & Requires empirical validation \\
\bottomrule
\end{tabularx}
\end{table}

\section{Post-hype evidence}

\subsection{Learning benefits are conditional rather than universal}

The strongest 2026 syntheses converge on a conditional picture.  A September 2026 review of 76 empirical studies in programming education likewise finds that benefits are most defensible when AI support is combined with adaptive scaffolding, process-oriented feedback, and human judgment, while inaccurate outputs, over-reliance, academic-integrity uncertainty, unequal access, and verification gaps remain recurrent implementation challenges \citep{katona2026}.  Kumar, Wongsirichot, and Nanthaamornphong's systematic review of 72 primary studies reports robust short-term efficiency gains, recurring over-reliance, weak transfer from AI-assisted completion to independent skill, and strong evidence for assessment redesign \citep{kumar2026}.  Adejumo and colleagues' systematic review of 64 empirical studies similarly emphasizes the dual role of GenAI: personalization and support can assist problem solving, while hallucinated or misleading outputs can increase cognitive burden and encourage over-reliance \citep{adejumo2026}.  Agbo and colleagues' broader computing-education review likewise finds expanding adoption across levels and subjects, but with both benefits and risks that depend on pedagogical strategy \citep{agbo2025}.  A 2026 review focused on software-development learning similarly identifies tutoring, feedback, assessment, code generation, computational thinking, and ethics as tightly coupled themes rather than independent uses of LLMs \citep{contreras2026}.

The recent STEM meta-analysis by \citet{boolzen2026} provides an especially useful caution against headline averages.  Its conventional random-effects model found a positive pooled effect on externally assessed cognitive outcomes, but heterogeneity was extreme ($I^2=96.32\%$) and the prediction interval included negative, null, and positive effects.  A robust Bayesian analysis that adjusted for publication bias favoured the absence of an overall positive or negative effect.  Knowledge-oriented outcomes performed differently from skill-oriented outcomes.  The implication for SE education is not that GenAI ``does not work.''  It is that average effects conceal design conditions and that procedural, evaluative, and professional skills cannot be inferred from short-term performance improvements.

Individual studies make the mechanism more concrete.  Structured prompt training can improve performance relative to untrained AI use \citep{garg2025}; controlled work also reports gains in computational thinking, programming self-efficacy, and motivation under a specific GenAI-supported intervention \citep{yilmaz2023}; guarded tutoring can provide on-demand help without directly revealing complete solutions \citep{liffiton2023}; and personalized Parsons-puzzle scaffolding can preserve engagement better than simply showing generated code \citep{hou2024}.  These studies support a narrow but important conclusion: access to a capable model is not a pedagogy.  The educational treatment is the combination of task design, permitted delegation, scaffolding, and the evidence students must subsequently provide.

\subsection{Productivity and completion are weak proxies for durable competence}

Generative systems make it unusually easy to confuse task completion with learning.  A student can submit working code yet be unable to trace a path, identify a hidden assumption, design an adequate oracle, explain a dependency, or adapt the code to a change request.  The distinction is visible in novice-programming research.  \citet{prather2024widening} observed that some students used GenAI as an accelerator while struggling students could experience compounded metacognitive difficulties and an illusion of competence.  \citet{qiao2026}, reviewing 31 studies of GenAI for code comprehension, found genuine opportunities for tailored explanation alongside inaccurate or unclear explanations and difficulties for novices in both prompting and evaluation.

The educational response should therefore sample competence after, around, and beyond the generating interaction.  Useful evidence includes an oral explanation, independent tests, code tracing, a delayed change request, defect localization, a design rationale, a critique of rejected AI output, or maintenance work under changed conditions.  These measures are not anti-AI.  They are ways to distinguish assistance from substitution.

\subsection{Assessment validity is a deeper problem than detection}

GenAI changes the evidentiary meaning of take-home artifacts.  Students and educators differ in how they interpret AI use and its impact on assessment \citep{zastudil2023,kizilcec2024}, and students encounter different rules across courses \citep{keuning2024,yabaku2024}.  A final artifact can contain substantial machine-produced code or prose while revealing little about the learner's reasoning.  In this setting, detection is an unstable organizing strategy: technically similar AI use can be pedagogically legitimate in one activity and disallowed in another.

Recent SE-specific evidence reinforces the importance of design and clarity.  In a 2026 FSE Companion study, \citet{santos2026} report that students' descriptions of inappropriate or disallowed LLM use clustered around programming assignments, routine coursework, and documentation, with time pressure and unclear expectations playing an important role.  The implication is not that misconduct disappears when policies are clearer; it is that assessment validity and integrity are partly design problems.  Courses need to define what assistance is permitted, what must be disclosed, and what evidence of independent competence remains required.

A stronger assessment architecture is therefore a portfolio rather than a single rule.  Classroom designs that make AI-generated artifacts objects of critique rather than hidden production shortcuts provide one concrete pattern for such redesign \citep{petrovska2024}.  Some tasks should be AI-restricted to establish foundational fluency.  Some should permit AI while requiring process evidence.  Some should require AI so students learn critique, verification, and governance.  Oral or live components can verify transfer.  Delayed modification can test whether understanding survives after the original interaction is unavailable.

\subsection{Prior knowledge and trust calibration matter}

Students are not homogeneous users.  \citet{shoufan2023} found that students valued ChatGPT's explanations while recognizing the need for background knowledge to judge them.  The ``widening gap'' study shows why this matters in practice: stronger or better-calibrated novices could ignore poor suggestions, while struggling students could be pulled further off course \citep{prather2024widening}.  Large reviews also identify over-reliance and prior knowledge as recurring moderators \citep{kumar2026,adejumo2026}.

This creates a calibration problem.  Fluent output can be most persuasive when the learner has the least independent knowledge with which to challenge it.  The solution is not simply better prompting.  Students need repeated encounters with output that is correct, obviously wrong, subtly wrong, locally plausible but contextually inappropriate, and correct for the wrong reason.  They also need independent adjudication mechanisms: tests, traces, specifications, documentation, measurements, threat models, and stakeholder evidence.  Calibrated reliance is a learning outcome in its own right.

\subsection{AI tutoring and feedback require deliberate design}

GenAI can provide scalable explanations and feedback, but the design of the interaction matters.  GPTutor explored code explanation inside a development environment \citep{chen2023gptutor}; CodeHelp explicitly constrained responses to avoid directly revealing solutions \citep{liffiton2023}; CodeTailor transformed students' incorrect code into personalized practice rather than returning a finished solution \citep{hou2024}.  These systems represent a shift from unrestricted chat to learning-oriented mediation.

Evidence about generated feedback also counsels restraint.  \citet{azaiz2024} found meaningful improvements in the structure and specificity of GPT-4 feedback on introductory programming submissions but also inconsistent feedback.  In a concurrent-programming course, \citet{estevez2025} found that general-purpose LLMs were not sufficiently accurate to substitute for expert assessment of concurrency errors.  AI can therefore extend first-line support, but instructors still need to decide what feedback can be automated, how students should respond to it, and where human diagnosis remains necessary.

\section{Students and faculty in the judgment-centred model}

A judgment-centred account is incomplete if it treats learners and instructors merely as endpoints of a tool intervention.  The same AI capability is experienced differently by a student seeking help, an instructor trying to infer learning, a team integrating generated work, and a program setting common rules.  The evidence therefore supports an actor-centred layer in the framework: student agency concerns how assistance is sought, interpreted, and owned; faculty judgment concerns how learning is designed, evidenced, and governed.

\subsection{Student perspective: help-seeking, authorship, and agency}

For many students, GenAI is becoming part of the ordinary ecology of help-seeking rather than a single, exceptional category of ``AI use.''  In an ACE 2024 study combining a survey of 47 computing students with eight interviews, \citet{houhelp2024} found that GenAI had not displaced traditional help resources; preferences varied with the task and with perceived quality, latency, and trustworthiness.  The study also reported preliminary evidence that effective AI help-seeking is itself a skill and that benefits may be unevenly distributed among students who differ in their ability to use LLMs productively.  A complementary 12-week deployment of an LLM-powered programming assistant with 52 students captured more than 2,500 queries: most sought immediate help with programming assignments, fewer pursued broader conceptual understanding, and students often supplied minimal contextual information \citep{sheese2024help}.  Together, these studies caution against treating access to an AI assistant as equivalent to effective help-seeking.

Student agency therefore involves more than prompt fluency.  Learners need to decide when AI is an appropriate source, when a human peer or instructor is preferable, how much context to provide, how to interrogate an answer, and when to stop asking for more generation and reason independently.  This interacts with prior knowledge.  Students value rapid explanations but recognize that background knowledge is needed to evaluate them \citep{shoufan2023}, while novice-programming research shows that poorly calibrated assistance can compound metacognitive difficulty for struggling learners \citep{prather2024widening}.  At the same time, specific structured interventions can support affective as well as cognitive outcomes: \citet{yilmaz2023} reported gains in programming self-efficacy and motivation under the intervention they studied.  These results should not be generalized into a universal motivational effect; they show why the form of support matters.

GenAI also makes authorship and ownership less binary.  A submitted artifact can combine a student's problem formulation, generated code, model explanations, generated tests, copied documentation, and later human edits.  Cross-course evidence shows substantial variation in how students use GenAI and in the rules they encounter \citep{keuning2024}, while student--educator studies document differences in how AI use and assessment are interpreted \citep{zastudil2023,kizilcec2024}.  Recent SE integrity research further links problematic or disallowed use to assessment conditions, time pressure, and unclear expectations rather than supporting a purely detection-centred response \citep{santos2026}.  The design implication is not to erase authorship boundaries, but to make the delegated work visible.  Policies should distinguish explanation, hints, code generation, test generation, documentation, review, and agent action, and should attach each permitted form of delegation to evidence that the learner still owns the engineering decision.

In this sense, student agency is not synonymous with unrestricted tool choice.  It is the capacity to seek assistance deliberately, calibrate reliance, preserve enough independent understanding to challenge output, disclose material delegation, and remain able to explain and change the resulting software.  Those capabilities connect the student perspective directly to AASEE's \obligation{explain}, \obligation{verify}, \obligation{modify}, and \obligation{account} obligations.

\subsection{Faculty perspective: assessment validity, capability, workload, and policy coherence}

Instructor caution is similarly better understood as an evidentiary problem than as resistance to technology.  When students can obtain plausible code, explanations, tests, and documentation externally, the final artifact carries less information about the cognitive process that produced it.  Interviews with computing students and instructors show both alignment and tension around appropriate GenAI integration \citep{zastudil2023}; a cross-national study of 680 students and 87 educators likewise found meaningful differences in perceived assessment impact and in preferred responses to GenAI \citep{kizilcec2024}.  Faculty therefore face a practical question: which combinations of artifacts, process evidence, oral explanation, independent work, and transfer tasks justify an inference that the intended learning occurred?

Recent discipline-specific qualitative evidence makes the institutional dimension clearer.  \citet{saravanan2026} conducted three focus groups with 16 engineering faculty at a U.S. research-intensive university.  Participants described a posture of reluctant engagement with GenAI, concerns about cognitive offloading and loss of productive struggle, disruption to established feedback and assessment practices, and frustration with gaps in coordinated institutional policy.  They also anticipated curricular redesign and emphasized the continuing value of experiential and mentorship-based learning.  These are faculty perceptions and experiences, not direct measurements of student learning, and the single-institution sample should not be generalized to all engineering faculty.  Their value is contextual: they show the kinds of pedagogical and organizational judgments instructors report having to make as AI capabilities enter routine coursework.

Responsible integration also changes faculty work.  Instructors may need to redesign tasks, decide which AI capabilities are compatible with an objective, evaluate prompts or other process evidence selectively, conduct oral or live checks, curate AI-generated feedback, protect student and client data, and keep policies current as tools change.  The literature does not yet provide a stable cross-institution estimate of this workload, so the issue should be treated as an implementation constraint rather than a quantified universal burden.  Nevertheless, recent engineering-faculty evidence explicitly reports pedagogical costs when policy gaps are left to individual instructors \citep{saravanan2026}, and SE-specific design work increasingly treats institutional support as part of the intervention.  Guide-AI-Ed links course and curriculum design choices to responsible GenAI use \citep{geruslu2026}; the maturity model proposed by \citet{alzahrani2026} addresses institutional readiness and progressive integration.

A judgment-centred program therefore needs coherence without imposing one universal classroom rule.  Programs can establish common principles for privacy, provenance, disclosure, prohibited data, high-risk uses, and evidence of understanding while allowing instructors to choose different AI conditions for different learning objectives.  Faculty development should likewise focus less on generic prompt tips and more on discipline-specific assessment patterns, failure modes, verification practices, and sustainable ways to inspect evidence.  The instructor's role is not simply to permit or police AI; it is to design the conditions under which delegated work remains educationally interpretable.

\section{Software engineering as a distinct educational problem}

A persistent weakness in broad computing-education discussions is the tendency to let programming stand in for software engineering.  The distinction matters.  Programming tasks can often be bounded around an algorithm or a small artifact.  SE artifacts are coupled to requirements, architecture, dependencies, test strategy, deployment, security, configuration, maintenance, teams, clients, and organizational processes.  The unit of responsibility is therefore larger than a code snippet.

This lifecycle coupling changes both the learning risk and the evidence needed to manage it.  In a capstone, AI-generated artifacts enter a shared codebase, persist across iterations, and may be delivered to an external stakeholder.  \citet{mircea2026} studied self-determined GenAI use in 18 real-world project teams and found use across lifecycle activities alongside practical concerns about verification, independent understanding, data protection, and team-level governance.  Complementing that capstone evidence, \citet{kharrufa2026} studied a semester-long second-year SE team project and found that GenAI's educational effects were entangled with the roles the tools played, the support students received, and the transparency of AI use to teammates as well as instructors.  These issues are hard to see in isolated programming exercises because the consequences arrive later: another developer must understand the code, a client changes a requirement, an integration fails, or a security assumption proves false.

Testing provides a particularly clear example.  In an FSE Companion experience report, \citet{yang2026} describe LLM use in software-testing education and highlight difficulties involving prompt context, constraints, and iterative strategy.  A 2026 observational study of undergraduate unit-testing work identified multiple interaction strategies, ranging from high AI reliance to primarily human ideation and implementation; participants reported time savings and support for test ideation, but also concerns about trust, quality, and ownership \citep{ardic2026}.  The key educational insight is that test generation is not test judgment.  Students still need to reason about equivalence classes, boundaries, oracles, mutation resistance, coverage, environment assumptions, and what a failure means.

Requirements and architecture expose similar limits.  An LLM can draft user stories or architectural alternatives, but requirements remain social, contextual, and value-laden, while architectural decisions depend on quality attributes, constraints, deployment environments, organizational capabilities, and future maintenance.  A strong AI-augmented assignment therefore treats generated material as a candidate to be evaluated against evidence rather than as an answer to be accepted.

This is why SE education should not isolate AI in a single ``prompt engineering'' module.  AI should appear where engineering judgment appears: requirements ambiguity analysis, design reviews, testing labs, code review, security assessment, configuration and deployment work, capstone governance, and maintenance.  The relevant competence is not model-specific prompt fluency.  It is disciplined delegation under lifecycle responsibility.

\section{From copilots to agents: the supervision problem}

\subsection{Agentic capability changes the authentic task}

The shift toward agentic software development is qualitatively different from faster autocomplete.  Repository-level benchmarks such as SWE-bench require models or systems to reason across real issues, multiple files, long contexts, and execution environments \citep{jimenez2024}.  Platforms such as OpenHands give software agents access to code editing, command-line tools, browsing, sandboxes, and evaluation tasks \citep{wang2025}.  Agentic SE research is beginning to describe engineering processes in which humans orchestrate goal-directed agents rather than directly authoring every change \citep{hassan2026}.

These developments should not be read as evidence that autonomous agents improve education.  They establish a future-facing authenticity problem.  If professional engineers increasingly delegate multi-step work to agents, then SE graduates need competencies that are poorly captured by ``write a better prompt'': task decomposition, specification of acceptance criteria, permission boundaries, environment isolation, provenance, test and evaluation harnesses, diff review, monitoring, stop conditions, rollback, incident recovery, and human accountability for accepted changes.

Computing-education researchers have begun to identify agentic AI itself as a distinct research challenge \citep{mak2026}.  Parallel conceptual work frames the educational target as an `agentic engineer' who specifies intent, orchestrates delegated work, verifies outcomes, and remains accountable for increasingly autonomous systems \citep{alenezi2026}.  Curriculum proposals are also emerging; for example, ASE-26 argues for treating agentic software engineering as a discipline with explicit attention to specification, verification, orchestration, and auditability \citep{gorsky2026}.  Such proposals are valuable signals of curricular experimentation, but they are not yet validated learning models.  AASEE therefore treats agentic work as one integration context rather than as the destination of every course.

\subsection{Agentic literacy is supervision literacy}

A useful educational distinction is between \emph{generation} and \emph{action}.  A conversational model can produce a candidate response that a student chooses whether to use.  An agent can make a sequence of changes before a human inspects every intermediate step.  The larger the action space, the more important constraints and recovery become.

For students, agent supervision should therefore be taught as ordinary SE responsibility expressed in a new workflow.  Requirements become task contracts and acceptance criteria.  Architecture constrains what an agent is allowed to change.  Testing becomes an independent evaluation harness rather than merely a generator of green checks.  Configuration management becomes branch isolation, diff review, provenance, and rollback.  Security becomes permission minimization, secret handling, sandboxing, and dependency review.  Project management becomes decomposition, delegation, monitoring, and escalation.  These are not replacements for SE fundamentals; they are reasons to teach them more deliberately.

A central assessment principle follows: \emph{agent-produced work should increase, not decrease, the amount of independently checkable evidence}.  If an agent modifies a repository, students should be able to explain the intended change, justify the permission boundary, inspect the diff, interpret test evidence, identify side effects, and describe a recovery plan.  The educational artifact is not the patch alone; it is the governed decision around the patch.

\section{Comprehension debt: from metaphor to educational risk}

\subsection{Definition and provenance}

Technical debt describes future cost created by expedient engineering choices that defer necessary work \citep{kruchten2012}.  \citet{ahmad2026} applies a related socio-cognitive idea to GenAI-assisted student software projects: \emph{comprehension debt}, the growing gap between what a team understands about its codebase and what it needs to understand to maintain and modify it.  Based on 621 reflective diaries from 207 students, the study identifies patterns including black-box acceptance, context mismatch, dependency-induced atrophy, and verification bypass, together with a mitigating pattern in which AI serves as a comprehension scaffold.

This review does not claim to originate that term.  Its contribution is to operationalize comprehension debt as an SE education and assessment lens.  We define it as the \emph{deferred learning, maintenance, and accountability cost incurred when AI-assisted artifact production outpaces the learner's or team's ability to explain, test, modify, and justify the artifact}.  This definition connects Ahmad's team-cognition account to observable educational evidence.

The construct is useful precisely because it is not an anti-AI claim.  Heavy AI use does not imply high comprehension debt.  A learner can delegate aggressively while maintaining low debt through active verification, explanation, integration, and maintenance.  Conversely, a single opaque but critical component can create high debt.  Frequency of AI use is therefore a poor proxy for the construct.

\subsection{Observable indicators}

At the individual level, candidate indicators include inability to explain control or data flow, weak prediction of behaviour, poor defect localization, inadequate test-oracle design, failure to identify edge cases, and inability to adapt an earlier AI-assisted artifact.  At the team level, indicators include uneven ownership, code that only one member can explain, undocumented provenance, superficial review, and repeated dependence on private AI interactions that other team members cannot reconstruct.

These indicators deliberately separate artifact quality from human understanding.  A polished artifact can coexist with high comprehension debt.  This is what makes the construct educationally important: conventional grading can reward output while leaving the future cost invisible.

\subsection{Assessment as a debt probe}

Many conventional SE activities can be repurposed as comprehension-debt probes.  A delayed change request tests whether understanding survives after the original AI interaction.  A code walkthrough exposes whether design rationale has been internalized.  Independent tests reveal whether a learner can challenge generated behaviour rather than merely execute it.  Live debugging, peer review, maintenance tasks, and cross-member explanation test transfer under changed conditions.

Measurement should therefore compare what was produced with what can later be demonstrated.  Candidate measures include explanation quality scored against a rubric, code-tracing accuracy, independently designed tests, defect localization, performance on delayed change requests, and consistency between documentation and observed behaviour.  At team level, review coverage, code-ownership distribution, and the ability of multiple members to explain a generated component can provide complementary evidence.

The construct remains a research agenda rather than a validated scalar metric.  Its boundaries also matter.  Comprehension debt is not identical to technical debt, academic misconduct, ordinary forgetting, or low code quality.  It specifically concerns a deferred understanding obligation created or amplified when artifact acquisition outpaces the cognition needed to maintain responsibility for the artifact.

\section{AASEE: a judgment-centred curriculum and assessment framework}

\subsection{Relation to recent frameworks}

AASEE is designed to complement, not replace, other emerging frameworks.  Kumar and colleagues' VIE framework emphasizes Verification, Implementation, and Equity across computing education \citep{kumar2026}.  Guide-AI-Ed provides a design-oriented model for responsible GenAI integration in SE education \citep{geruslu2026}.  Alzahrani and Alghamdi's maturity model focuses on institutional readiness and progressive process areas \citep{alzahrani2026}.  ASE-26 proposes a dedicated curriculum for agentic software engineering \citep{gorsky2026}.  These contributions answer different questions.

AASEE's narrower purpose is to connect the \emph{scope of delegated engineering work} and its \emph{lifecycle consequence} to \emph{observable evidence of retained human competence}.  It is not an institutional maturity model and does not assume that higher levels are inherently better.  A first-year tracing exercise may appropriately remain AI-restricted even in a program that teaches agent orchestration elsewhere.  Conversely, a capstone can require agent supervision while still protecting independent checkpoints.  The level is selected from the learning objective, not pursued as a ladder.

\subsection{Four design propositions}

Four propositions summarize the framework's design logic.

\begin{description}[style=nextline,font=\normalfont\bfseries]
\item[P1 --- Delegation must not outrun verification.] The more work an AI system performs, the stronger the learner's independent means of checking that work must become.  Convenience is not evidence.
\item[P2 --- Generated artifacts require transfer evidence.] As direct authorship becomes less informative, assessment should increasingly sample explanation, debugging, maintenance, and adaptation under changed conditions.
\item[P3 --- Responsibility expands with lifecycle coupling.] AI use in a disposable exercise has different consequences from AI use in a shared repository, client project, or deployed system.  Governance should scale accordingly.
\item[P4 --- Independent fluency remains a prerequisite for judgment.] Protected practice is justified when the learning objective is the mental model that later makes AI critique possible.  Authenticity does not require automation in every learning episode.
\end{description}

\subsection{Five integration levels}

Table~\ref{tab:levels} summarizes the five AASEE integration levels, their educational focus, typical forms of delegation, and the evidence or safeguards associated with each.

\begin{table}[t]
\centering
\caption{Five non-linear AASEE integration levels.  A course selects a level from its learning objective; the levels are not a maturity ladder.}
\label{tab:levels}
\footnotesize
\begin{tabularx}{\textwidth}{P{0.18\textwidth}Y Y Y}
\toprule
\textbf{Level} & \textbf{Educational focus} & \textbf{Typical delegation} & \textbf{Required evidence / safeguards} \\
\midrule
1. Tool awareness & Capabilities, limitations, policy, provenance, privacy, failure modes, calibrated trust & AI output is mainly material for comparison and critique & Protected no-AI baselines; failure analysis; source and provenance checks \\
2. Guardrailed AI-assisted development & Bounded help while building foundations & Hints, explanations, debugging suggestions, examples, small fragments & Independent predictions; staged help; prompt/process disclosure; explanation checkpoints \\
3. AI-collaborative engineering & AI participates in substantive lifecycle work & Requirements drafts, design alternatives, implementation, refactoring, review, tests, documentation & Design rationale; independent tests; peer review; walkthroughs; delayed modification \\
4. AI governance & Team and client responsibility & Governed use of assistants across shared projects & Tool/data policy; provenance; review ownership; privacy rules; disclosure and escalation paths \\
5. Agentic and AI systems engineering & Supervision of multi-step agents and AI-enabled systems & Repository agents, tool use, evaluation loops, AI components & Task contracts; permission boundaries; sandboxes; evaluation harnesses; monitoring; rollback/recovery; security review; human accountability \\
\bottomrule
\end{tabularx}
\end{table}

Level 1, \emph{tool awareness}, builds calibrated trust.  Students learn capabilities, limitations, provenance, privacy, hallucination and failure modes, and local policy.  AI output is primarily an object of critique.  This level is fully compatible with protected no-AI exercises that establish baseline fluency.

Level 2, \emph{guardrailed AI-assisted development}, permits constrained help such as hints, explanations, debugging suggestions, examples, or small code fragments.  Guarded tutoring and scaffolded support motivate this level \citep{liffiton2023,hou2024}.  The design preserves productive struggle through staged help, independent predictions, and checkpoints where the learner must reason without generation.

Level 3, \emph{AI-collaborative engineering}, allows AI to participate across substantive engineering activities: requirements drafts, design alternatives, implementation, refactoring, code review, test generation, and documentation.  Students compare alternatives and remain responsible for integration.  Evidence shifts toward tests, design rationale, peer review, code walkthroughs, and delayed modification.  The central risk is false ownership: a team can appear productive while shared understanding fragments.

Level 4, \emph{AI governance}, moves responsibility to team and client contexts.  Teams define permitted tools, prohibited data, provenance rules, review responsibilities, disclosure norms, and escalation paths.  This level treats integrity as professional governance rather than only a classroom rule.  Capstone evidence makes these concerns concrete \citep{mircea2026}.

Level 5, \emph{agentic and AI systems engineering}, covers systems in which AI components or software agents can take multi-step actions.  Learning outcomes include task decomposition, permission boundaries, environment isolation, evaluation harnesses, monitoring, rollback, security review, and accountability for agent-produced changes.  This level is motivated by professional and technical capability trends \citep{jimenez2024,wang2025,hassan2026}, not by an assumption that autonomous development has already been pedagogically validated.

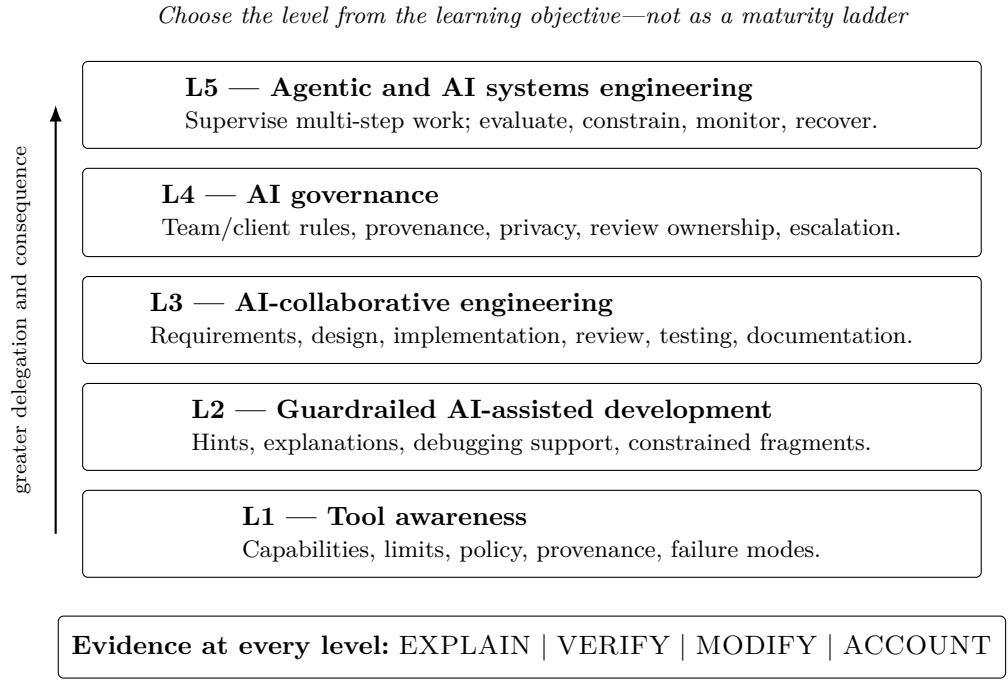
\begin{figure}[t]
\centering
\begin{tikzpicture}[font=\small,
  box/.style={draw, rounded corners=2pt, minimum width=.72\textwidth, minimum height=0.83cm, align=left, inner sep=6pt},
  arrow/.style={-{Latex[length=2.2mm]}, thick}]
\node[box] (l5) {\textbf{L5 --- Agentic and AI systems engineering}\\\footnotesize Supervise multi-step work; evaluate, constrain, monitor, recover.};
\node[box, below=2.5mm of l5] (l4) {\textbf{L4 --- AI governance}\\\footnotesize Team/client rules, provenance, privacy, review ownership, escalation.};
\node[box, below=2.5mm of l4] (l3) {\textbf{L3 --- AI-collaborative engineering}\\\footnotesize Requirements, design, implementation, review, testing, documentation.};
\node[box, below=2.5mm of l3] (l2) {\textbf{L2 --- Guardrailed AI-assisted development}\\\footnotesize Hints, explanations, debugging support, constrained fragments.};
\node[box, below=2.5mm of l2] (l1) {\textbf{L1 --- Tool awareness}\\\footnotesize Capabilities, limits, policy, provenance, failure modes.};
\node[draw, rounded corners=2pt, below=5mm of l1, minimum width=.72\textwidth, minimum height=.82cm, align=center, inner sep=5pt] (ev) {\textbf{Evidence at every level:} \obligation{EXPLAIN} $\mid$ \obligation{VERIFY} $\mid$ \obligation{MODIFY} $\mid$ \obligation{ACCOUNT}};
\draw[arrow] ($(l1.west)+(-0.35,0)$) -- ($(l5.west)+(-0.35,0)$);
\node[rotate=90,anchor=south,font=\scriptsize] at ($(l3.west)+(-0.55,0)$) {greater delegation and consequence};
\node[above=3mm of l5,font=\footnotesize\itshape] {Choose the level from the learning objective---not as a maturity ladder};
\end{tikzpicture}
\caption{The AASEE framework. Increasing delegation changes the kind of evidence and governance required, but higher levels are not inherently superior.}
\label{fig:aasee}
\end{figure}

Figure~\ref{fig:aasee} summarizes the five integration levels together with increasing delegation and consequence and the four evidence obligations that apply across all levels.

\subsection{Four cross-cutting evidence obligations}

The framework makes learning assessable through four obligations.

\obligation{Explain} means the learner can articulate what an artifact does, why key decisions were made, and what assumptions matter.  Explanation should be specific enough to support prediction and review, not merely paraphrase documentation.

\obligation{Verify} means claims are challenged with tests, static or dynamic analysis, traces, measurements, specifications, domain evidence, or stakeholder validation.  Verification should be sufficiently independent that a model is not merely asked to grade its own work.

\obligation{Modify} means understanding transfers to a changed requirement, defect, environment, or constraint without simply replaying the original interaction.  Modification is especially valuable because it reveals comprehension debt that polished first submissions can hide.

\obligation{Account} means the learner can document provenance, disclose material AI assistance, protect data, state review responsibility, and accept responsibility for the final engineering decision.  At higher levels, account also includes permissions, monitoring, escalation, and recovery.

Together, these obligations shift assessment from visible authorship to visible judgment.  They also avoid a common category error: assuming that more sophisticated tools imply more sophisticated pedagogy.  A state-of-the-art model can be used at Level 2 if the task is tightly bounded; an ordinary assistant can create Level 4 governance obligations if its output enters a client workflow.

\subsection{The capability--risk--safeguard rule}

An instructor can operationalize AASEE with a simple sequence: state the learning objective without reference to a tool; identify the AI capability that could be used; name the specific learning or engineering risk created by that capability; select evidence that makes retained human competence visible; then choose the integration level and assessment conditions appropriate to the consequence of failure.

If AI generates code, the risk may be solution substitution and weak ownership; safeguards can include independent tests, explanation, and a later change request.  If AI drafts requirements, the risk may be generic or incomplete stakeholder assumptions; safeguards can include ambiguity analysis and stakeholder validation.  If an agent edits a repository, the risk includes unintended multi-file changes and poorly bounded action; safeguards include constrained permissions, a reviewable diff, an independent evaluation harness, and a recovery plan.

Table~\ref{tab:matrix} illustrates this capability--risk--safeguard rule across representative AI capabilities and uses.

\begin{table}[t]
\centering
\caption{Examples of the capability--risk--safeguard rule.}
\label{tab:matrix}
\footnotesize
\begin{tabularx}{\textwidth}{P{0.18\textwidth}Y Y Y}
\toprule
\textbf{AI capability} & \textbf{Potential use} & \textbf{Learning / engineering risk} & \textbf{Visible safeguard} \\
\midrule
Explanation / tutoring & Concept help, debugging guidance & False mastery; persuasive but wrong explanation & Predict first; compare explanations; trace behaviour; restate and defend \\
Artifact generation & Code, tests, docs, design alternatives & Substitution; shallow authorship; over-reliance & Independent tests; rationale; change request; provenance \\
Review / critique & Code review, risk identification & Automation bias; missed deeper defects & Human peer review; targeted instructor sampling; defect seeding \\
Requirements simulation & Mock users, user stories, acceptance criteria & Generic, biased, or fabricated stakeholder model & Validate against real stakeholders/domain evidence; identify assumptions \\
Workflow automation & Summaries, repetitive edits, migration & Deskilling; hidden dependency changes & Before/after analysis; regression tests; dependency review \\
Repository agents & Multi-file change, test execution, tool use & Unbounded action; side effects; opaque provenance & Task contract; least privilege; sandbox; diff review; evaluation harness; rollback \\
\bottomrule
\end{tabularx}
\end{table}

\section{Curriculum and assessment implementation}

\subsection{Course-design procedure}

AASEE can be implemented without rebuilding every course around AI.  An instructor begins with the competence students must possess at the end of an activity.  The tool decision comes second.  This ordering prevents novelty-driven assignments whose AI component is clearer than their learning purpose.

A testing course, for example, may have the objective of reasoning about adequacy, fault detection, and oracles.  A Level 3 activity can require students to obtain an initial test suite from an LLM.  The generated suite is only the starting artifact.  Students classify the input space, identify missing boundaries, justify each oracle, execute the tests, add cases that target uncovered behaviour, and later respond to a seeded defect or change request.  The grade attaches to analysis and transfer, not to the number of tests produced.  This design aligns with the prompt-context and workflow difficulties observed in testing education \citep{yang2026,ardic2026}.

A senior capstone can combine Levels 4 and 5 without turning the project into an autonomy demonstration.  Before an agent touches a repository, the team defines a task boundary, permitted data, tool permissions, branch or sandbox constraints, acceptance tests, review ownership, and a recovery plan.  Agent-produced changes enter ordinary engineering review: the team inspects the diff, runs independent tests, checks architecture and security consequences, records material AI involvement, and ensures that more than one member can explain the change.  In client-facing work, the team also decides what information may be shared with external models.  The educational artifact is the team's governed engineering decision, not simply the agent's patch.

\subsection{Assessment portfolios rather than universal policies}

No single institution-wide rule such as ``AI allowed'' or ``AI banned'' can serve all learning outcomes.  Programs need assessment portfolios.  Protected tasks establish independent fluency.  AI-permitted tasks teach disciplined use with transparent process evidence.  AI-required tasks teach critique, evaluation, and governance.  Oral and live components verify transfer.  Maintenance tasks and delayed changes test whether understanding persists.

Useful patterns include generate-then-verify assignments, AI-assisted testing followed by oracle analysis, human-versus-AI code review comparisons, prompt-as-specification exercises, requirements ambiguity analysis, oral defences, and comprehension-debt audits.  The common property is that the grade is attached to evidence of judgment rather than to the mere presence or absence of AI.

\subsection{Program-level progression}

The framework is most useful when applied across a program.  Early courses can combine Level 1 awareness with protected foundations and bounded Level 2 help.  Intermediate courses can move into Level 3 through testing, refactoring, review, requirements, and design exercises.  Advanced project courses can make Level 4 governance an explicit project-management responsibility.  Courses in advanced SE, DevOps, security, intelligent systems, or capstones can introduce Level 5 supervision of agents and AI-enabled components.

This progression avoids two symmetric failures: graduates who can prompt but cannot reason independently, and graduates who know fundamentals but have never learned to govern AI-mediated workflows.  AI literacy should be embedded into SE fundamentals rather than added as a detached prompt-engineering unit.  Prompt precision becomes requirements precision; verification becomes testing; model critique becomes review; provenance becomes configuration and documentation discipline; disclosure and escalation become professional accountability.

\section{Research agenda for durable competence}

The next phase of research should move beyond adoption surveys and short demonstrations.  Several questions deserve priority.

\textbf{Delayed transfer and maintenance.}  Studies should measure what students can explain, debug, test, and modify after the original AI interaction is unavailable.  Multi-course and longitudinal designs are needed to determine whether AI scaffolding builds competence or merely improves immediate output.

\textbf{Comprehension debt.}  The field needs validated instruments that distinguish artifact quality from human understanding.  Candidate measures include explanation quality, tracing accuracy, independent test design, maintenance performance, and cross-member team understanding.  Construct validity and predictive value should be tested across institutions and course levels.

\textbf{Scaffolding withdrawal.}  If guardrails are useful, when and how should they be removed?  Research should compare static support with progressive withdrawal, testing whether students become more independent rather than merely more comfortable with the tool.

\textbf{Equity and calibration.}  Prior preparation, confidence, language background, access, and disability may change who benefits from AI support.  The goal is not simply equal tool access but equitable development of verification and judgment capabilities.  Recent reviews identify equity and prior knowledge as underdeveloped dimensions \citep{kumar2026,adejumo2026}.

\textbf{Team cognition.}  Team-project evidence already shows that GenAI use can reshape roles, support patterns, and transparency inside student teams \citep{kharrufa2026}.  SE research should now examine who understands AI-produced changes, how review responsibility is distributed, how private AI conversations affect shared mental models, and whether governance practices survive under deadline pressure.

\textbf{Agent supervision.}  There is not yet a mature evidence base on learning with repository agents.  Studies should separate agent productivity from educational outcomes and evaluate task decomposition, permission design, evaluation quality, review behaviour, recovery, and security reasoning.  Agentic environments also create new experimental variables: autonomy, tool access, memory, execution privileges, and human intervention points.

\textbf{Faculty workload and institutional infrastructure.}  Responsible integration changes the work of educators.  Studies should measure the cost of evaluating process evidence, administering oral defences, curating AI-generated feedback, maintaining tool policy, and supporting privacy and accessibility.  A pedagogically sound intervention that is operationally unsustainable will not scale.

\textbf{Framework evaluation.}  AASEE itself should be treated as an evidence-informed design hypothesis.  Comparative studies can test whether its explain--verify--modify--account obligations improve assessment validity, reduce comprehension debt, or improve calibration relative to equally permissive AI access without those obligations.  The five integration levels should be revised if empirical work reveals different boundaries or missing responsibilities.

\section{Limitations and threats to validity}

This review has five important limitations.  First, it is integrative rather than systematic.  The original review process did not retain a complete database-by-database screening trail, so this revision cannot support prevalence estimates or a PRISMA claim.  Recent systematic reviews are used as coverage anchors, but relevant work may still be missing.

Second, the primary literature is heterogeneous.  Models, interfaces, courses, populations, tasks, durations, outcome measures, and permitted forms of AI assistance differ substantially.  The 2026 meta-analysis shows that this heterogeneity is not a minor statistical nuisance; it can reverse how pooled results should be interpreted \citep{boolzen2026}.

Third, many studies measure perception, immediate performance, or short-term task outcomes.  Durable understanding, maintenance ability, professional judgment, and cross-course transfer remain less certain.  Findings about student enthusiasm or productivity should not be generalized to long-term competence.

Fourth, some of the most SE-specific evidence is new.  Capstone and comprehension-debt work from 2026 is highly relevant but still early \citep{mircea2026,ahmad2026}.  The agentic literature is even less mature educationally.  Agent benchmarks and platforms establish capability, while curriculum proposals and research agendas establish questions; they do not demonstrate that particular agentic pedagogies improve learning.

Fifth, the framework is normative.  Its propositions combine evidence with an educational objective: graduates should remain capable of taking responsibility for software outcomes.  AASEE is therefore not presented as a validated causal model.  Its value depends on empirical testing, refinement, and comparison with alternative frameworks.

\section{Conclusion}

Generative AI has made software production easier to separate from software understanding, and agentic AI is beginning to separate action from continuous human authorship.  For software engineering education, that separation is the central design problem.  The best current evidence does not support either universal prohibition or unstructured adoption.  It supports a conditional view: AI can provide explanations, feedback, practice, and productivity benefits, but learning depends on task design, prior knowledge, scaffolding, verification, and what is subsequently assessed.  Recent meta-analytic evidence also warns against treating positive short-term results as a universal learning effect.

Software engineering widens the problem beyond programming.  Generated artifacts enter architectures, teams, repositories, test pipelines, client relationships, and maintenance histories.  As agents gain the ability to act across tools and files, students must learn not only how to request work but how to bound, evaluate, monitor, reject, and recover from delegated action.

The judgment-centred approach proposed here treats AI policy as a curriculum-design decision rather than an ideological stance.  AASEE's five levels range from tool awareness through guardrailed assistance and collaborative engineering to team governance and agentic/AI systems engineering.  Its four cross-cutting obligations---\obligation{explain}, \obligation{verify}, \obligation{modify}, and \obligation{account}---specify what human competence must remain visible as delegation increases.  Comprehension debt provides a complementary diagnostic: when students or teams can produce more than they can later explain, test, maintain, or justify, the curriculum has hidden rather than solved the learning problem.

The educational goal is therefore not to maximize either manual production or AI use.  It is to develop engineers who know what to delegate, can determine whether the result is trustworthy, can recover when it is not, and remain accountable for the software that reaches users.  In an era of increasingly capable generative and agentic systems, those are not secondary ``AI skills.''  They are becoming central expressions of software engineering judgment.

\section*{Data and materials}
This study reports a synthesis of publicly available publications and does not collect new human-participant data.  Bibliographic identifiers and version-of-record references are provided in the reference list.  Because the original narrative-review process did not retain a complete screening log, no claim of exhaustive retrieval is made.

\bibliographystyle{plainnat}
\bibliography{references}

\end{document}